%% file: main.tex
\documentclass[nonacm,sigconf,authorversion,balance]{acmart}

\usepackage{booktabs} 
\usepackage{balance}

\setcopyright{none}

\usepackage{algorithmic}
\usepackage{graphicx}
\usepackage{textcomp}
\usepackage{listings}
\usepackage{pgfplots}
\pgfplotsset{width=7cm,compat=1.8}
\graphicspath{ {images/} }
\usepackage{multicol}
\usepackage[T1]{fontenc}
\usepackage{verbatim}

\usepackage[most]{tcolorbox}

\definecolor{editcolor}{rgb}{0, 0, 0}
\newcommand*{\edit}{\color{editcolor}}

\usepackage{stfloats}
\usepackage{mathrsfs}
\usepackage{float}
\usepackage{url}
\usepackage{graphicx}

\usepackage{multirow}
\usepackage{xcolor}
\usepackage{booktabs,xcolor,siunitx}

\usepackage{array}

\usepackage{threeparttable}

\usepackage{pifont}

\usepackage{adjustbox}

\usepackage{hyperref}

\usepackage{upgreek}

\usepackage{tabularx,ragged2e,booktabs,caption}
\usepackage{tikz}

\newcommand\copyrighttextbottom{%
\footnotesize Accepted version of Macak, M., Stovcik, M., Rebok, T., Ge, M., Rossi, B., Buhnova, B. CopAS: A Big Data Forensic Analytics System. Proceedings of the 8th International Conference on Internet of Things, Big Data and Security (IoTBDS), 2023, SciTePress.}
\newcommand\copyrightnoticebottom{%
\begin{tikzpicture}[remember picture,overlay]
\node[anchor=south,yshift=10pt] at (current page.south) {\fbox{\parbox{\dimexpr1.0\textwidth-\fboxsep-\fboxrule\relax}{\copyrighttextbottom}}};
\end{tikzpicture}%
}

\begin{document}
\title{CopAS: A Big Data Forensic Analytics System}
  
\renewcommand{\shorttitle}{CopAS: A Big Data Forensic Analytics System}

\author{Martin Macak}
\affiliation{%
  \institution{Faculty of Informatics\\Masaryk University}
  \city{Brno} 
  \country{Czechia} 
}
\email{macak@mail.muni.cz}

\author{Matus Stovcik}
\affiliation{%
  \institution{Faculty of Informatics\\Masaryk University}
  \city{Brno} 
  \country{Czechia} 
}
\email{mstovcik@mail.muni.cz}

\author{Tomas Rebok}
\affiliation{%
  \institution{Institute of Computer Science\\Masaryk University}
  \city{Brno} 
  \country{Czechia} 
}
\email{xrebok@fi.muni.cz}

\author{Mouzhi Ge}
\affiliation{%
  \institution{Deggendorf Institute of\\Technology (DIT)}
  \city{Deggendorf} 
  \country{Germany} 
}
\email{mouzhi.ge@th-deg.de}

\author{Bruno Rossi}
\affiliation{%
  \institution{Faculty of Informatics\\Masaryk University}
  \city{Brno} 
  \country{Czechia} 
}
\email{brossi@mail.muni.cz}

\author{Barbora Buhnova}
\affiliation{%
  \institution{Faculty of Informatics\\Masaryk University}
  \city{Brno} 
  \country{Czechia} 
}
\email{buhnova@mail.muni.cz}

\renewcommand{\shortauthors}{Macak et al.}

\begin{abstract}
With the advancing digitization of our society, network security has become one of the critical concerns for most organizations. In this paper, we present CopAS, a system targeted at Big Data forensics analysis, allowing network operators to comfortably analyze and correlate large amounts of network data to get insights about potentially malicious and suspicious events. We demonstrate the practical usage of CopAS for insider attack detection on a publicly available PCAP dataset and show how the system can be used to detect insiders hiding their malicious activity in the large amounts of data streams generated during the operations of an organization within the network.

\end{abstract}

%
%
%

\keywords{Network Security, Network Traffic Analysis, Forensics Analysis, Big Data, Insider Attack Detection}

\maketitle

\copyrightnoticebottom

\section{Introduction}

Insider attacks are one of the most significant cybersecurity issues in organizations~\cite{insiderSurveyHomoliak,RUU}.
Their impacts include financial loss, disruption to the organization, loss of reputation, and long-term impacts on organizational culture~\cite{impacts}, which makes them important to study. Since insiders are authorized employees with access to the organization’s resources and the knowledge of its internal processes, their attacks are significantly more challenging to detect than external ones~\cite{aresPaper,insiderSurveyHong}. 

Existing solutions for detecting insider attacks show that Big Data involved in the analysis is a major challenge~\cite{insiderSurveyGheyas,insiderSurveyLiu}. It often relies on analyzing large volumes of data (e.g., network traffic) over a long time span, making the analysis very time-consuming and challenging.
This is especially true when network traffic data captured in PCAP files are analyzed on a per-packet basis using tools like Wireshark or similar applications. An alternative to these per-packet analyses is an analysis at the level of individual network flows. Here, the set of packets belonging to a single network connection is described by a single network flow record with appropriate descriptive information (e.g., source and destination IP address, source and destination port, protocols used, amount of data transferred, and timestamps). Analyzing entire network flows based on their descriptive information is thus much more comfortable from the forensic analyst's point of view, allowing them to gain so-called situational awareness more easily.

Since the extraction of network flows from PCAP files requires some preprocessing (e.g., correcting any problems in the PCAP files, merging for connections captured in multiple files, appropriate flow extraction and description, and indexing in an appropriate database), this paper presents the CopAS system that we have developed intending to ameliorate and simplify this process for effective support of police investigation.
CopAS combines a set of existing tools with several handy features and a user-friendly graphical interface, allowing the analyst to focus on the data analysis itself rather than on the preparation and configuration of the infrastructure and preprocessing configuration, making it a unique tool for complex, more effective and straightforward network captures analysis. 

In this paper, we present CopAS architecture and features and demonstrate its usefulness for detecting insider cyberattacks in an organization's network. We also examine where to draw the line between the automated preparation of the analysis and the analysis that the investigator wants better control over.

We provide the following main contributions in this paper:
\begin{enumerate}
    \item The provision of the CopAS platform for the support of digital forensic analysis integrating and combining several tools for Big Data network analysis. The platform is free to use for any interested party\footnote{Publicly available at \url{https://gitlab.ics.muni.cz/bigdata/CopAS}};
    \item Demonstration of the application of CopAS to support forensic analysis for insider attack detection by using the CSE-CIC-IDS2018 dataset~\cite{datasetPaper};
\end{enumerate}

The remainder of the paper is structured as follows. Section~\ref{sec2} provides an overview of work on insider attack detection and related Big Data platforms. In Section~\ref{sec3}, our CopAS platform is introduced and described. Section~\ref{sec4} demonstrates CopAS in detecting the insider attack, followed by the discussion in Section~\ref{sec5}. Section~\ref{sec6} concludes the paper.

\section{Related Work}
\label{sec2}

Two main directions that are relevant to our work are network-based insider attack detection approaches and platforms for digital forensic analysis.

Approaches that use a network-based detection of insider attacks are Lv et al. \cite{lv2019modeling} and Kholidy et al.~\cite{kholidy2020}, which reuse a dataset proposed in previous research in Kholidy et al.~\cite{kholidy2012}. Other approaches use host-based analysis, for example, MS Word commands~\cite{elMasri}, OS activities~\cite{insiderSurveySalem}, audit logs~\cite{Macak2020}, and UNIX commands~\cite{YuandGraham,KIM2005160}. However, the practical usage of network traffic Big Data for insider attack detection remains an unexplored challenge~\cite{insiderSurveyGheyas,insiderSurveyLiu}.

Over time, many platforms for digital forensic analysis {\edit (Table \ref{table:relplatforms})} emerged to provide support for Big Data Analysis and provide ways to integrate and link knowledge to support police investigation and security events  \cite{marciani2017,schroeder2007automated}. The needs of such platforms are mainly to integrate a plethora of tools/systems available (such as Pig, Hadoop, Cassandra, Zookeeper, Lucene, and Mahout) and different types of analysis required for big digital forensics analysis, such as link analysis to connect knowledge from different sources (e.g., \cite{marciani2017,schroeder2007automated}) or text/data mining approaches supported by machine learning~\cite{pramanik2017big}.

CrimeLink Analysis Explorer \cite{schroeder2007automated} is a platform that provides support for link analysis investigations, supporting co‐occurrence analysis, the shortest path algorithm, and a heuristic to identify the importance of associations. The platform was developed as an ad-hoc solution based on a management system supported by a database connection and modules for co-occurrence weights, a heuristic module, an association path module, and a graphical user interface. However, it was not meant to scale over Big Data but rather to look into the benefits of having a platform for knowledge integration.

Another platform for digital forensics analysis was proposed by \cite{marciani2017}. It is a data stream processing platform based on the Apache Flink Big Data framework, Apache Kafka for event processing, and Neo4J for data storage and visualization. The experimental evaluation has shown that the platform was effective for criminal link analysis, reaching an accuracy of 82\% in linking different sources.

CrimeNet Explorer \cite{xu2005CrimeNet} is a~framework for automated criminal network analysis and visualization. It allows to build, analyze, and visualize crime networks based on communication between involved entities. The platform is based on social network creation from crime databases, clustering of nodes, structural analysis, and visualization of network partitions. 

Other platforms for packet inspections were developed on top of existing forensics network analysis tools. One example is XplicoAlerts \cite{gacimartindetecting}, built on top of the Xplico tool \cite{costa2012xplico}, combining packet inspection and browsing to filter and detect potential attacks. The idea behind XplicoAlerts is to support the analysis by automatic alerts when the network traffic contains suspicious communications worth further investigation. XplicoAlerts provides an interface to analyze and annotate suspicious events, allowing a user to get an aggregated view for large-scale analysis of network data events.

The platforms proposed by \cite{kumar2013scalable} and \cite{lee2010internet} are examples of platforms based on Hadoop and MapReduce to provide scalable intrusion detection platforms. Such platforms are based on network traffic log parsers, storage, and analysis in Hadoop/MapReduce, with analyzed logs provided for further analysis/filtering of suspicious communications.

As different from these existing platforms, the CopAS platform proposed in this paper can address the specific detection needs by means of support of network traffic analysis with the integration of well-known frameworks (e.g., \textit{ElasticSearch}, \textit{Kibana}), allowing the person involved in forensic analysis to have a single platform, in which all the data analysis is integrated. Compared to the discussed platforms, containerization supports high flexibility as well as large-scale data analysis, allowing the analyzer to adjust the needs based on the amount of data available.  This is especially relevant for network traffic data analysis since the data to be analyzed can grow exponentially based on the number of users involved. Additionally, further extensibility of the platform can be developed by the integration of other frameworks depending on the needs for data analysis that arise.

\begin{table}[htb]
\centering
\renewcommand{\arraystretch}{1.8}
\caption{Platforms for digital forensics analysis}
\label{table:relplatforms}

\begin{tabular}{ |p{0.5cm}|p{2.4cm}|p{4.5cm}| } 
 \hline
 \bf{Year} & \bf{Platform} & \bf{Focus}  \\ 
 \hline 
 2017 & N/A \cite{marciani2017} & Big Data link analysis investigations\\
 2013 & N/A \cite{kumar2013scalable} & Suspicious network traffic analysis\\
 2012 & XplicoAlerts \cite{gacimartindetecting} & Crime investigation of network traffic\\
 2010 & N/A \cite{lee2010internet} & Suspicious network traffic analysis\\ 
 2007 & CrimeLink Analysis Explorer~\cite{schroeder2007automated}  & Link analysis investigations\\
 2005 & CrimeNet Explorer \cite{xu2005CrimeNet} & Criminal Network Analysis and visualization\\
 \hline
\end{tabular}

\end{table}

\section{CopAS System}
\label{sec3}

This section introduces the CopAS platform with a detailed overview of its essential context, requirements, architecture, and implementation.

\subsection{Application Context}

When dealing with cyberattacks and (digital) crime investigations, network traffic captures are highly-valued data allowing the analyst to understand the situation faced. However, an analysis of network traffic captures -- usually encapsulated in packet captures (PCAP format) -- is a very exhaustive and time-consuming process since it is very complicated for a data analyst to build awareness of the captured situation on the level of individual IP packets. Moreover, this process becomes even more complicated and often even impossible when dealing with large amounts of such captures. Thus, it is highly beneficial to preprocess such packet captures and extract higher-level information, such as compound information about all the individual network flows, which is more easily understandable by humans and keeps all the necessary information required by network data analysts. However, such a transformation is not the only preprocessing step required to be done in order to index network captures in a powerful analysis system. During the pre-analysis phase, it is often required to enrich these data in various manners (like resolved DNS names, geographic information related to IP addresses, etc.) as well as to maintain various unpredictable states in order to make the transformation successful (like fixing various errors that may occur in packet captures).

To make the depicted complex process of preprocessing network traffic captures and their analysis more effective and straightforward, we proposed and developed a solution that employs the Elastic framework~\cite{elastic} and facilitates this process in a user-friendly manner. The solution, called CopAS (the acronym stands for \emph{Cop's Analytical System}), combines a set of existing tools with a user-friendly graphical interface, allowing the network data analysts to focus just on the data analysis itself, not on the technical process of packet captures' preprocessing and indexing.

Regarding the CopAS analytical features, we precisely selected a set of integrated analytical tools so that the CopAS can be used for an analysis of various cybersecurity attacks/incidents. The insider attack detection, which we analyze later in this paper, requires the analytical tools to provide the analyst with a list of all the captured network data flows enriched with information like IP addresses and/or DNS resolved names of communicating parties, port numbers, and amounts of data transmitted -- and all of these enriched with timestamps, making the detection of the sequence of the attack events possible. Moreover, the flow description should be enriched with the detected network protocols and their header information, including the data payloads transmitted in open form. The structure of these data nicely fits into the model of document-oriented databases (like ElasticSearch, MongoDB, and others), allowing the analysts to query them using complex queries. These queries help to identify insider attacks, which are often complex and complicated. However, since such complex attacks can be hard to read by a human analyst, a robust visualization framework such as Kibana or Arkime/Moloch is also necessary to better understand the query results and gain awareness about the captured situation. We thus decided to integrate these tools into the CopAS to provide the analysts with sufficient flexibility and high analytical features.

\subsection{System Requirements}

When designing CopAS, we have taken the following major requirements into account:

\begin{itemize}
\item \emph{Data analysis features} -- besides streamlining the whole preprocessing phase for network traffic analysis, the tool has to be extensible for different data formats (like JSONs or CSVs) as well as for different analysis tools (like Arkime, formerly known as Moloch), allowing to get different views on the analyzed situation.

\item \emph{User-friendliness} -- as already mentioned, usability is the crucial aspect of the CopAS solution. Since the depicted process of network captures analysis requires several steps to be performed, and that requires an adequate level of IT knowledge, we want the tool to make the whole process easier, allowing the analysts to focus just on the data analysis itself, not on the technical details.

\item \emph{Performance} -- since the amount of network captures can be very large, the tool has to provide a sufficient level of performance, effectively using the hardware infrastructure that it is deployed on. Thus, it has to fine-tune all the individual tools as well as the employed data workflow and introduce as low additional overhead as necessary. Moreover, scalability is an important factor that has to be addressed as well.

\item \emph{Flexibility} -- despite hiding unnecessary technical details behind a user-friendly interface, CopAS has to allow flexible process modifications (individual steps configuration) during the preprocessing phase. Moreover, on the hardware resources level, the solution has to allow its users to use available hardware resources flexibly: e.g., smaller network captures could be analyzed on less powerful personal computers, and once the amount of data or analyses rises, it should allow its users to migrate the analysis to more powerful servers and continue their analysis.

\item \emph{Automation} -- since the preprocessing and indexing tasks are often repeatable -- for example, new network captures are preprocessed and analyzed in the same way and with the same process configuration as the previous ones -- the tool has to support automation of such repeatable tasks as much as possible.

\item \emph{Data and System isolation} -- usually, there is a need to analyze data from multiple cases, sometimes with the need to adapt the configuration of integrated tools or even extend them. We decided to isolate the entire stack of analytical tools in each container to support these use cases, also featuring the possibility of renewing a fresh container state once the configuration becomes misbehaving.

\item \emph{Ease of deployment} -- even though being part of user-friendliness, we explicitly wanted the tool to be easy and user-friendly to deploy, no matter what hardware resources or operating system it is deployed on.

\item {\edit \emph{Ease of analysis} -- to further support the ease of its practical deployment and use, CopAS should support the integration/import of pre-made user analytical dashboards (e.g., sets of graphs/tables for analysis of known attacks or statistical information). Such dashboards will allow the data analyst to quickly obtain basic information about the captured situation or, for example, will enable her to identify known attacks more quickly.}
\end{itemize}

\subsection{CopAS Architecture}

To address all the primary requirements, we decided to build the CopAS from individual widely-used components, suitably integrated into a unified complex solution. While the required data analysis features are provided by the set of integrated tools (currently Elastic framework and Arkime), the user-friendliness of data preprocessing/indexing and automation is achieved by a web-based interface we developed. Once used by a user, the interface properly configures (for the sake of performance) and starts all the underlying tools necessary to provide the required processing, gather their outputs, and adapt its further stages to them. To support flexibility and ease of deployment features, we decided to employ containers -- lightweight virtual machines that allow us to make the CopAS independent of the underlying operating system and provide features for the flexible creation and migration of various analytical projects.

\begin{figure}[h]
\begin{center}
    \includegraphics[width=0.45\textwidth]{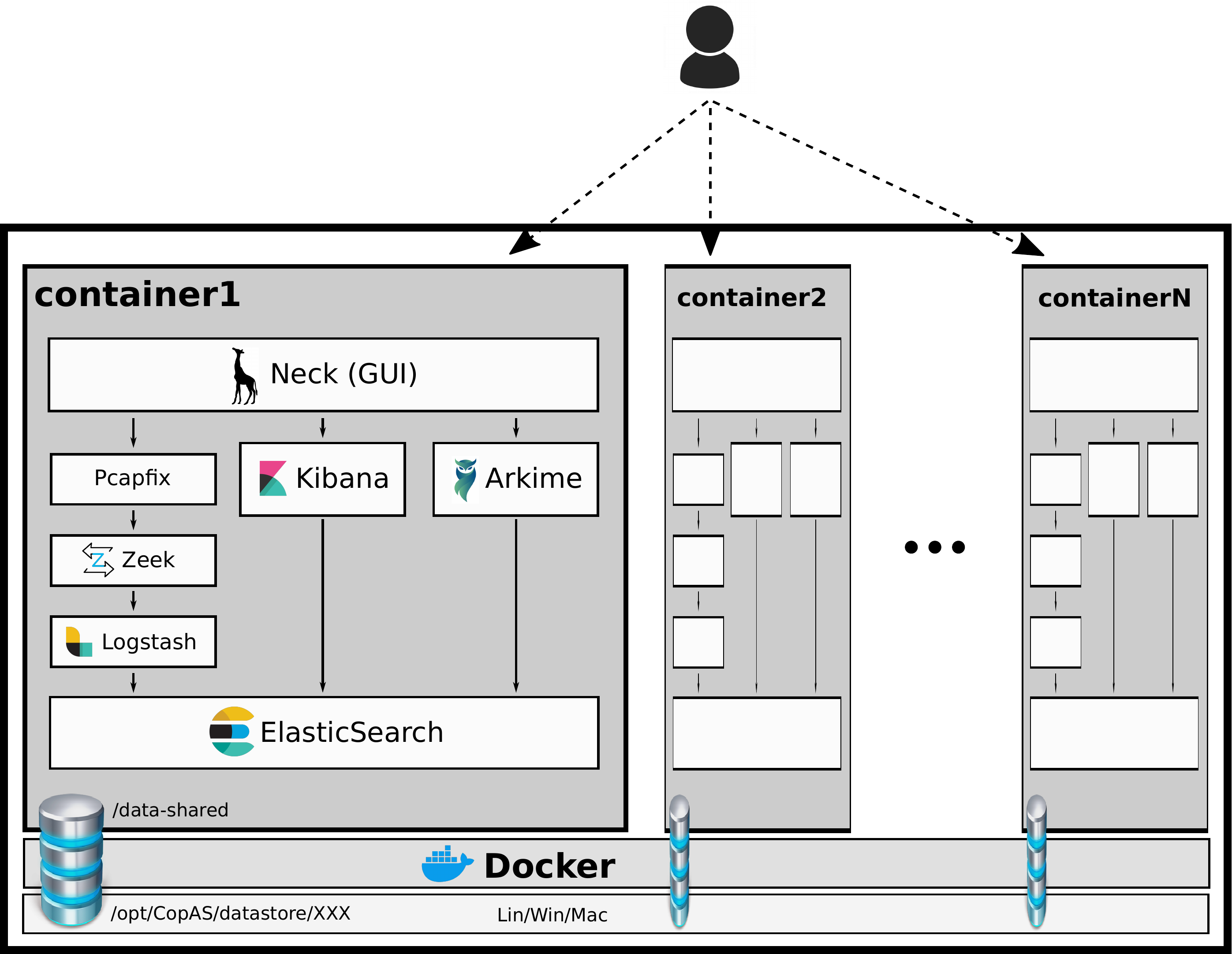}
    \caption{CopAS architecture schema, illustrating a set of containers with integrated analytical tools and provided user interface.}
    \label{fig:copas-architecture}
\end{center}
\end{figure}

As the CopAS architecture depicted in Figure~\ref{fig:copas-architecture} shows, all the necessary preprocessing and analysis tools are encapsulated in containers, which are then individually managed on the host(s). This architecture allows users to perform various data analysis projects simultaneously -- each encapsulated in a particular container -- and flexibly switch among them. Moreover, the individual containers (individual analytical projects) can be created or destroyed, started or stopped, backed up or restored, or even flexibly moved/migrated across various CopAS host instances (e.g., from less powerful hardware to more powerful ones).

The current CopAS implementation employs Docker~\cite{docker} as the container management engine, controlled by a shell script allowing to perform various operations on the container level. Inside each CopAS container, there is a full stack of analytical tools required for performing analyses of network captures or other input data, including the graphical user interface, which allows configuring and running all the preprocessing phases and analytical tools in a user-friendly fashion. Currently, the CopAS containers employ the following essential set of tools:

\begin{itemize}
\item \emph{Neck} -- a web-based graphical interface we developed to make the process of data preprocessing and analysis easier. Neck is a web-based solution that allows CopAS users to manage local files stored in the container/host (upload, download, move, delete, etc.), choose the ones for further analysis, and perform the whole data preprocessing phase in a user-friendly fashion. During several steps, users can adapt the default configurations (if required), as well as store frequently used configurations of the whole preprocessing phase for further use. Finally, based on the input data format, Neck instructs and starts the necessary tools to prepare chosen data for uploading into the ElasticSearch database (or alternative integrated tools, like Arkime) for further analysis.

\item \emph{ElasticSearch} -- an open-source NoSQL search engine developed by Elastic NV company. ElasticSearch is the essential part and heart of CopAS, allowing it to index all the analyzed data and provide an interface for querying them in (near) real-time for interactive analyses in Kibana. 

\item \emph{Kibana} -- an open-source analytical interface, also developed by Elastic NV company. Kibana provides the primary interface for data analysts, allowing them to specify data queries (in the background sent to ElasticSearch), visualize their results, and thus iteratively and interactively build their awareness about the situation captured inside the analyzed dataset.

\item \emph{LogStash} -- an open-source tool employed from the toolset developed by Elastic NV company. LogStash serves CopAS for conversion and transformation of input data files, enhancing them (e.g., DNS resolving), and finally uploading them into the ElasticSearch database.

\item \emph{Zeek} (previously known as \emph{Bro}) -- an open-source network monitoring and analysis framework that processes IP packets captured in PCAP format. CopAS uses Zeek to process input packet captures and transforms them into network flows described by several attributes (like originator and responder IP addresses, timestamp of connection establishment, amount of data sent, and network protocols used), which are further forwarded to ElasticSearch and indexed.

\item \emph{Arkime} (formerly known as \emph{Moloch}) -- an open-source, large-scale engine for indexing and searching network packet captures, serving as an alternative approach to ElasticSearch/Kibana data analyses, demonstrating the extensibility of CopAS. While ElasticSearch/Kibana data analysis could be considered more generic, Arkime is a highly-specialized tool for network security analysts, providing a set of specific features that can make their analysis more efficient.

\item \emph{pcapfix} -- as its name suggests, pcapfix is an open-source tool able to check for and repair various errors inside (corrupted) network traffic captures in PCAP format. Even though those captures are usually without errors, CopAS uses pcapfix for safety reasons so that the processed data can be considered consistent and error-free.
\end{itemize}

In addition to the graphical user interface streamlining the preprocessing phase performed by the described fine-tuned and properly configured toolset, CopAS implements various additional handy features that make its practical use more comfortable and efficient for data analysts. Those features incorporate, for example:

\begin{itemize}
\item established shared directories between each container and its hosting system, that make transfers of data files between the host system and particular containers easier,
\item possibility of indexing and further analysis of input data in various formats (currently supported formats are Packet Captures -- PCAP, JavaScript Object Notation -- JSON, and Comma-Separated Values -- CSV), which makes CopAS a generic data indexing and analysis tool (not only specialized on network captures),
\item ability to directly work with various data compression archives (currently ZIP and TGZ), which does not require the data analysts to extract them on their own,
\item availability of so-called \emph{CopAS WatchDogs}, which periodically look over specified directories for new data files and automatically index them using user-defined configuration,
\item ability to interconnect several CopAS containers into a single distributed system, allowing to index and analyse huge datasets on a set of physical computers/servers,
\item possibility to enter each container from command line, allowing the user to adapt its (system) configuration and/or integrated tools,
\item {\edit  the ability to export/import created analytical dashboards from Kibana so that the data analyst can re-use them in another analysis to gain the situation awareness faster,}
\item (\emph{running implementation}) graph model-based analyses of (network) data, which will allow the analysts to perform more efficient analyses of various complex relationships among individual entities and their communication. The implementation is based on the Dgraph graph database,
\item (\emph{running implementation}) support for drive image captures (in the IMG file format), allowing forensic analysts to index and analyze filesystem structure and files' timestamps of hard drives.
\end{itemize}

\subsection{Implementation}
As mentioned previously in the paper, CopAS is implemented as a set of suitably selected tools together with their integration and automation of some aspects -- a web-based graphical interface that guides the user through all the necessary steps to properly index required data.
Once the CopAS is installed, the user is given a command-line utility that allows manipulating data-analytic containers. The utility can create a new container, start and stop it, backup or load (i.e., migrate), enter into its command line, or even destroy it. Besides these container functions, the utility also provides a set of functions for showing relevant information about running containers, monitoring their resources, updating the base container image, or providing necessary debugging information.


The creation and the complete start of a newly created container usually takes a few seconds (tens of seconds at most): once started, the user is provided with an URL address with its port number and thus unique for each created container, where the container's web-based user interface listens behind. The CopAS main user interface then provides the user with a set of functions that are usually performed in the following order:

\begin{itemize}
    \item \emph{File Manager} -- a simple web-based file manager that allows the user to upload, manipulate and destroy data inside the analytical container. While this provides a simple and intuitive way of preparing the data necessary to analyze, an alternative approach of uploading through a hosting system and a unique directory shared between the host and the container, which is useful, especially for large datasets, is also provided. 
    \item \emph{Import} -- a step-by-step import function that indexes the chosen data and prepares them for further analysis. While the function tries to choose the proper setting for the detected data automatically (e.g., PCAPs vs. CSVs), it also allows the user to variously adapt the setting of all the individual steps (e.g., Logstash service configuration) in a user-friendly way. During the indexing configuration, the user can choose a set of directories that will be monitored for new data uploads (for their automatic import) and choose whether to upload the data into the Elastic framework or the Arkime/Moloch (or both).
    \item \emph{Kibana} and \emph{Arkime/Moloch} -- functionality that forwards the user to the graphical interfaces of these integrated tools.
    \item \emph{History} -- a list of performed analyses showing the list of indexed directories and the particular configurations.
    \item \emph{Elastic Status}, \emph{ElasticSearch Cleanup}, and \emph{Arkime/Moloch Cleanup} -- a set of service functions that are useful for checking the status of the Elastic database subsystem (used for both Kibana and Arkime/Moloch), as well as for the ElasticSearch and Arkime/Moloch cleanup.
\end{itemize}

The implementation and configuration of all the tools are realized for maximum performance and optimum resource usage (automatically detected and adapted based on the host system's resources). For example, besides proper configuration of the individual tools, the web-based interface transparently to the user combines the detected flows of multiple input files into a single large data stream, thus minimizing the overhead of starting all the individual analytical tools, shortening the time necessary for data import.

\begin{tcolorbox}[colback=gray!5!white,colframe=gray!75!black,title=CopAS Main Highlights]
       CopAS is containerized platform that allows the scalability of digital forensic analysis based on network traffic by supporting all the phases of data analysis: from data pre-processing, data cleaning to data visualization.
\end{tcolorbox}

\begin{figure*}[h]
    \centering
    \includegraphics[width=0.99\textwidth]{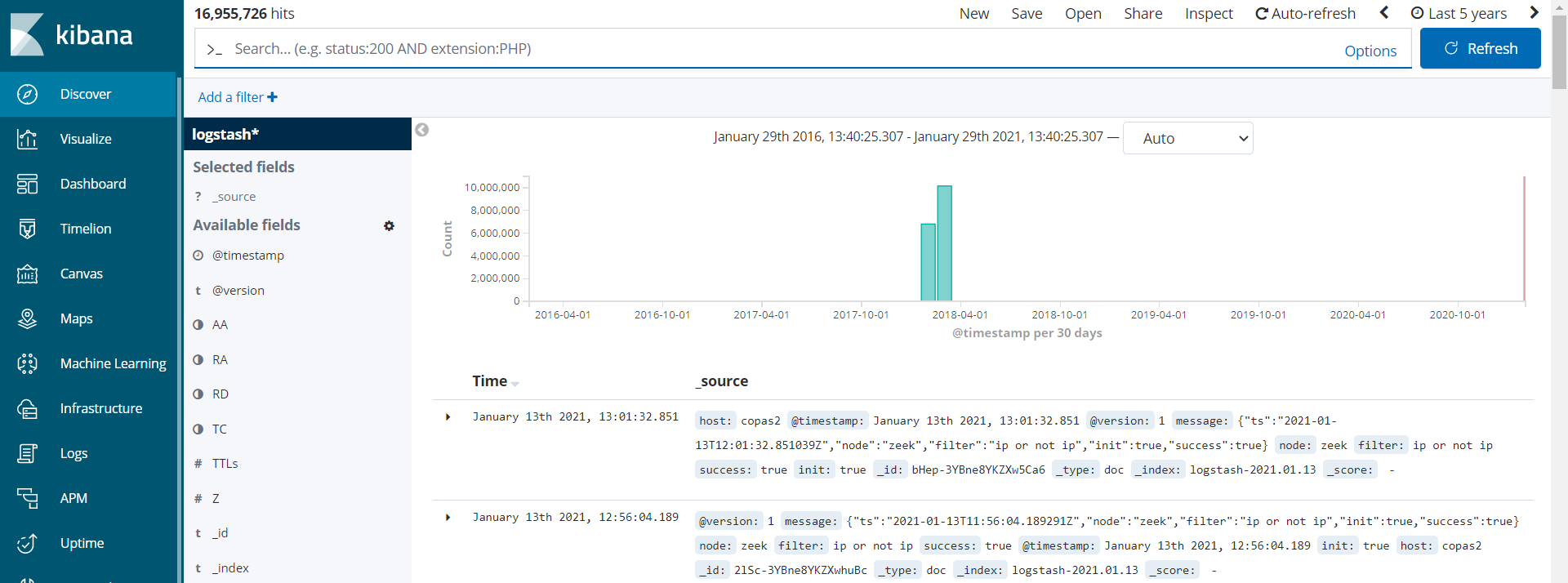}
    \caption{CopAS Kibana Dashboard Integration.}
    \label{fig:copas-kibana-dashboard}
\end{figure*}

\subsection{Analysis features}

\begin{figure*}[h]
    \centering
    \includegraphics[width=0.99\textwidth]{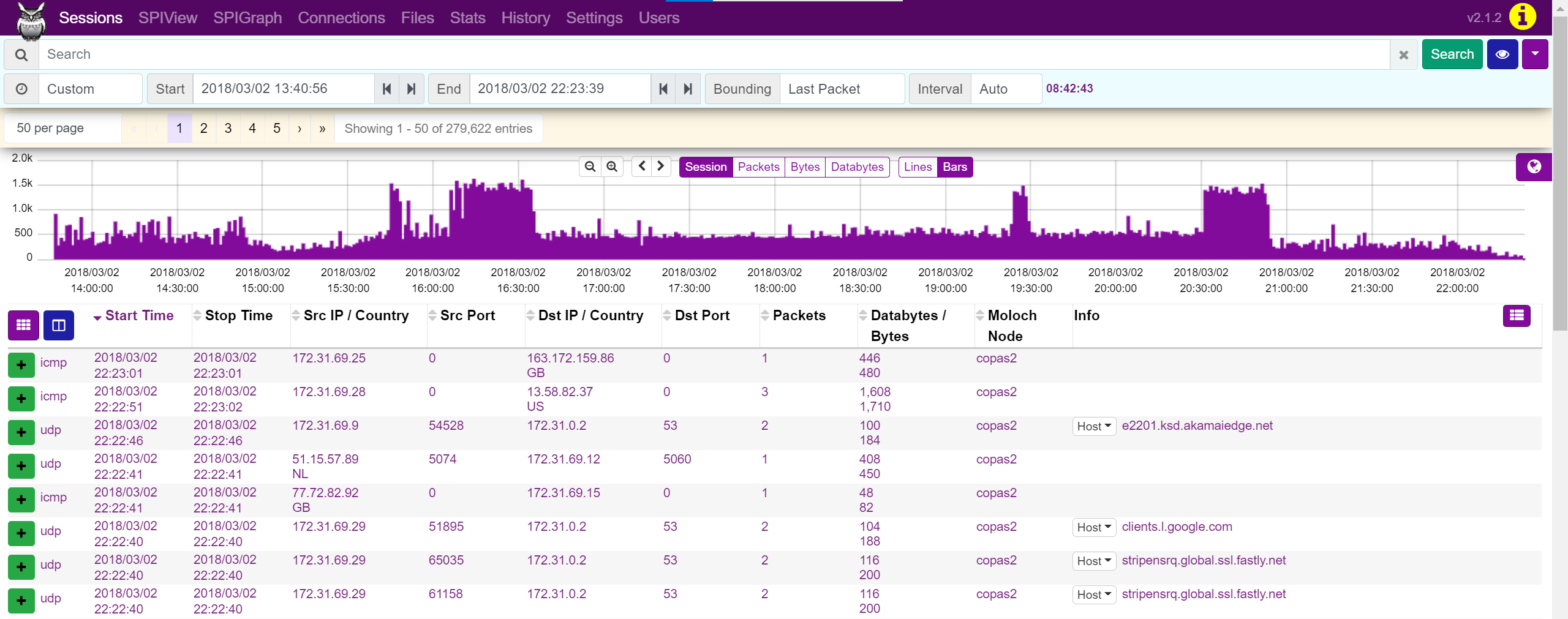}
    \caption{CopAS Arkime/Moloch Dashboard Integration.}
    \label{fig:copas-moloch-dashboard}
\end{figure*}

Once the data are properly indexed, the processes of mining and crime/incident discovery will take place. While the CopAS primarily focuses on simplifying and shortening the processes of toolset configuration and proper dataset preparation  and indexation, its analytical features are intentionally provided by the integrated and widely-used analytical toolset. Currently, CopAS integrates two tools available to the analyst for data analysis:

\begin{itemize}
    \item \emph{Kibana} -- a widely used analytical tool that provides a generic query language and various visualization possibilities, allowing the user to visualize and analyze responses to the provided analytical queries interactively. In the CopAS, the Kibana serves both for the analysis of network traffic captures as well as for interactive analysis and visualization of various datasets (indexed as large CSV files). Its dashboard is shown in Figure~\ref{fig:copas-kibana-dashboard}.
    \item \emph{Arkime/Moloch} -- a highly specialized tool for digital forensics, providing a set of features focused on the analysis of network traffic captures. Even though not being as generic as Kibana is, the Arkime/Moloch enriches the CopAS features with a fine-tuned analytical interface, e.g., for digital forensics analysis and connection/communication graphs. Its dashboard is shown in Figure~\ref{fig:copas-moloch-dashboard}.
\end{itemize}

In the case of indexing the network traffic captures, all the individual packet captures are automatically transformed into an indexed set of detected network flows. All these flows are described by a set of their attributes -- like initiator's/responder's IP addresses and port numbers, (optionally) their DNS names and geographical locations, flow timestamps and duration, amount of data and packets transferred, and detected protocols -- that are available for their filtering, aggregations, and visualizations, supporting the process of building situation awareness.

\section{\uppercase{Experiment with CopAS}}
\label{sec4}
In this section, we demonstrate the capabilities of CopAS in supporting forensic analysis for insider detection. CopAS provides features that are useful for  data analysis on large datasets, such as the deployment and indexing of datasets with custom parameters that can be overridden. By using the analytical tool provided within this system, we emphasize the importance of CopAS in the context of the PCAP dataset analysis. To demonstrate its capabilities, we focus on a port scan attack as a result of an intrusion exploit by using a publicly available dataset~\cite{datasetPaper}. In this case, the external attacker gains access to the organization's network and afterward begins the insider attack, masquerading as their victim inside the organization.

\subsection{Goal}

The goal of the CopAS demonstration is to search for suspicious behavior of network participants; more closely, we look for insider port scan attacks. 
Unusual or abnormal load on ports between participants in the network can be an indicator of a serious attack. In regular case scenarios, participants use a relatively small number of ports between each other. Having a record and detailed information about abnormal port usage can be a significant help in the detection of insider attacks.
A port scan is an attack that scans a network for vulnerabilities. These vulnerabilities may lead to exploiting a known vulnerability of that service~\cite{portScanCit}. By detecting this attack, we can prevent unauthorized access to the devices in an organization. In the demonstration, we are interested in detecting which devices in the private network behave with malevolent intent, utilizing unusual amounts of unique ports. 

\subsection{Experimental Setting}
In order to thoroughly examine the chosen dataset, we use the proposed CopAS tool. CopAS Docker is hosted on a machine with Ubuntu 20.04.1 LTS (Focal Fossa) operating system. Our hosting machine is based on an Intel Core i7-4790K and 16 GiB memory. However, CopAS is not limited in any way to the usage of resources; it can scale to operate on the largest set of resources we can offer. 

We are using the CSE-CIC-IDS2018 \cite{datasetPaper} dataset on AWS\footnote{https://registry.opendata.aws/cse-cic-ids2018/}. {\edit This dataset consists of different attacks executed on the implemented infrastructure.} For each day, there is a specific attack. We index all days in the dataset. The whole dataset has around 17 million records and 66,741 unique IP addresses participating in the network. To showcase the usage of CopAS, we sampled the whole dataset with two days of port scan attacks occurring on two days: 28.2. and 1.3. As described by the authors of the dataset~\cite{datasetPaper}, we consider three subjects within an attack. The first subject, an attacker, attacked the network by sending a malicious program or exploiting a known backdoor. The second subject, an insider, which is the victim of an outside attacker, was infected by the mentioned malicious content and unwillingly forced to perform an attack on a private network, in our case, a port scan attack. The third subject, a victim within a private network, was affected by being the victim of a port scan attack performed by an insider~\cite{datasetPaper}. 

\subsection{Description of analysis}
We first used the CopAS platform for indexing our PCAP dataset. For analysis of indexed data, we use a built-in tool within CopAS, Kibana. Kibana offers different ways to approach this problem. We can use visualizations, for example. A visualization in Kibana is relatively easy to use --- with the ability to aggregate the data by giving visual feedback to the user. In our demonstration, we use the Kibana console, a tool offering enhanced functionality. The aim is to flag infiltrated or infected devices within a private network. Results from the analysis would lead to the physical checking of flagged devices.  

\begin{figure}[h]
\begin{center}
    \includegraphics[scale=0.6]{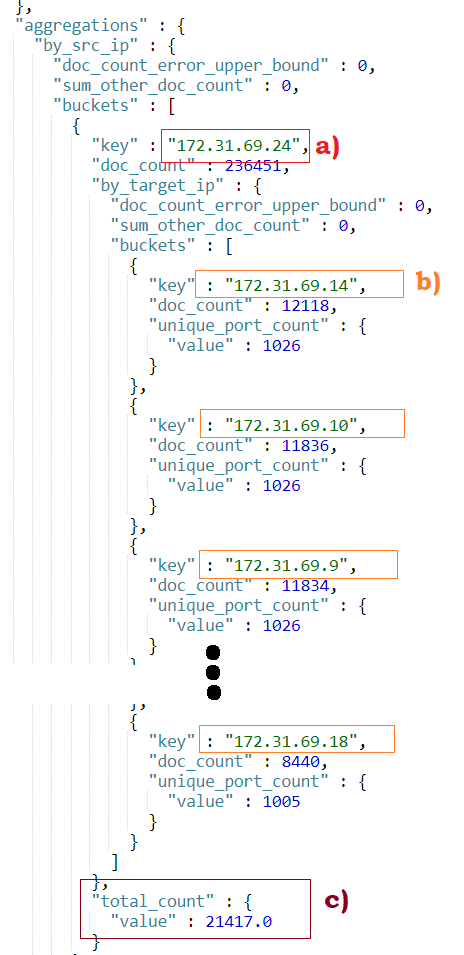}
\caption{Query result example}
\label{fig:demonstration}
\end{center}
\end{figure}

We implemented a solution that returns a result in JSON format (Figure \ref{fig:demonstration}), from which we can say report which IP addresses had performed a port scan attack. The result consists of buckets; each bucket is defined by one IP address -- shown in the figure as a). Within this bucket, there is
a list of every IP address -- shown in the figure as b) with a count of unique ports as "value"; we consider only values higher than ten. There is a considerable number of frequently used ports higher than ten. We set this threshold based on the inspection of the dataset; however, such a threshold can be customized, taking into account the properties of the dataset analyzed by considering the statistical distribution of historical data collected. Each value represents how many unique ports were used by IP defining the bucket. Last but not least, we have "total\_count" -- shown in the figure as c); this number represents the sum of before mentioned values. Only IP addresses with a total count higher than 500 are included in the results. Also, in this case, we based the selection of this threshold on the data distribution.  Based on the total of these counts for every IP, we can deduce the threshold from which behavior is considered suspicious and may be malevolent. 
For the demonstration, we have to increase the number of max buckets within our system. The default value of 10\,000 is not enough to correctly compute the results.

For our purpose, we use filters, sorting, and aggregation functions.
Our main metric is the number of unique ports used between each pair of IPs within our dataset. The query is designed to create buckets. For each IP in the dataset, we have one bucket. This upper-level bucket consists of other lower-level buckets representing every IP that received packets from an upper-level bucket. The lower-level bucket stores the unique count of used ports. We sum up a unique count of used ports for each upper-level bucket and store this value in the variable \emph{total\_count}.
We introduce some filters to sieve our data of IP with an irrelevant number of used ports for getting more precise results. In order to obtain the top results, we use the bucket sort functions within Kibana to sort in descending order.

This aggregated setup  consists of a list of all possibly infiltrated IPs, ordered by the sum of all the unique ports that the device used. 
Based on the results, we deduce which IPs in a network are infected by observing a significant difference in port usage between IPs.

\subsection{Experimental Results}

In this section, we present our results in the form of various charts. The results are filtered on thresholds determined within the analysis and given in the queries presented before. Therefore some columns in the chart are rounded to zero.

\begin{figure}[h]
\begin{center}
    \includegraphics[scale=0.46]{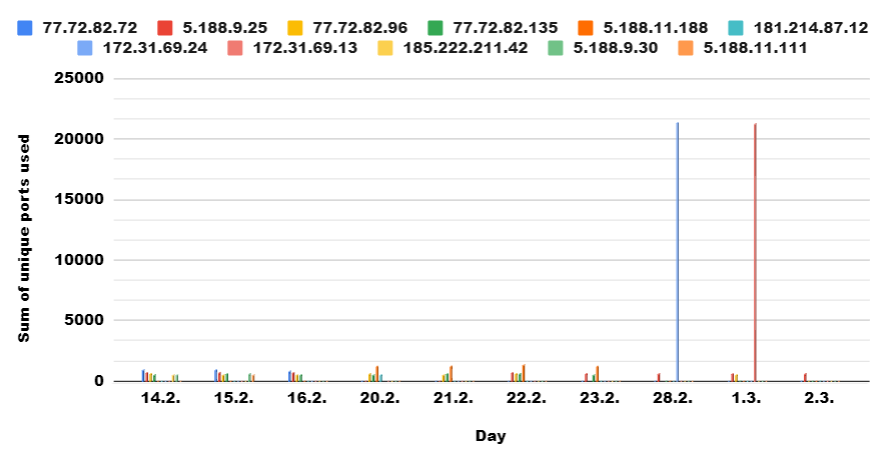}
\end{center}
\caption{The sum of unique ports used per a sender IP address for each day}
\label{fig:eachDayCount}
\end{figure}

We first provide a description of our query and introduce our method to filter unrelated outcomes. We only consider those pairs of IP addresses in which more than ten ports are used. Then we proceed even further by limiting our results. For each IP address, we have the sum of its used unique ports. We filter this sum on conditions higher than five hundred. 
Based on this approach, we got the results shown in Figure \ref{fig:eachDayCount}.

In Figure \ref{fig:eachDayCount}, we can see that having the sum of used unique ports around 1000 is common and found in six cases. The graph shows two abnormal values produced by IP addresses 172.31.69.24 and 172.31.69.13. Those two addresses were considered performing the port scan attack by the authors of the dataset and were identified by the analysis with CopAS.

\begin{figure}[h]
\begin{center}
    \includegraphics[scale=0.45]{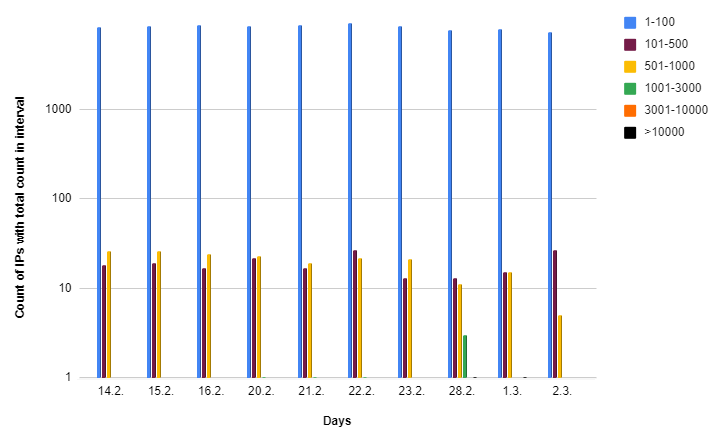}
\end{center}
\caption{The count of IP addresses with total count in intervals per day (logarithmic scale)}
\label{fig:eachDayInterval}
\end{figure}

Figure \ref{fig:eachDayInterval} is based on a slightly modified query from what is seen previously in Figure \ref{fig:eachDayCount}. In this case, we consider those pairs of IP addresses in which at least one port is used. Instead of looking for IP addresses with an abnormal amount of unique ports used in sending packets, we look at how many IP addresses have this amount within given intervals. Results show that most IP addresses use at most 100 ports per day, with some exceptions when this number rises to 3000. The graph also captured the abnormal values of IP addresses performing port scan attacks, as detected by CopAS.

This demonstration serves as an example of what CopAS is capable of. The demonstration is based on the presumption that in the private network of an organization, entities are not expected to perform any malicious behavior towards each other as they are one unit. Therefore by discovering malicious behavior between respective entities within the company, we can say there is a possibility of an insider attack.
Based on the presented results and method of discovering malicious behavior (port scan attack) within an organization, analysts can infer the problem and confirm the possibility of such an attack.

\begin{tcolorbox}[colback=gray!5!white,colframe=gray!75!black,title=CopAS Experimentation]
       The application of CopAS for network insider attack detection has shown how the platform's integration with several tools allows for easy data inspection by operators, pinpointing potentially malicious events and permitting the operators to perform additional inspections.
\end{tcolorbox}

\section{\uppercase{Discussions}}
\label{sec5}

While the captured network data can be analyzed using Wireshark or similar tools, we present a solution that supports the analysis at the level of individual network flows. Here, the set of packets belonging to a single network connection is described by a single network flow record with appropriate descriptive information used for the analysis. Analyzing entire network flows is thus much more comfortable from the forensic analyst's point of view, allowing her to gain so-called situational awareness more easily.

CopAS is not a yet-another analytical tool; instead, it introduces and implements a uniform and easy-to-use analytical environment integrating existing, highly specialized, and properly selected analytical tools to streamline this process. It thus serves as a 'glue' of these analytical tools, making the infrastructure setup, data import, and necessary data pre-processing tasks faster and more comfortable. As its features are inspired by the feedback gained from real-life investigations of police data analytics, it integrates a set of other handy features -- like the ability to integrate user analytical dashboards, isolation of different investigated cases, support for analysis of other file types, etc.

To demonstrate its usefulness and readiness for real-life analytical scenarios, we have used the CopAS platform for insider attack detection, looking at insider port scan attacks. In particular, we used CopAS support for network traffic analysis with the integration of ElasticSearch and Kibana. After indexing the sample PCAP dataset, we used visualizations in Kibana by aggregating and filtering data to look into unique port scans to cluster potentially infiltrated devices based on unique ports used. We defined thresholds based on the historical distribution of the data to identify suspicious devices. Utilizing CopAS architecture based on containers, we could take a snapshot of the situation on the days considered for the demonstration. More instances of analysis could have been started in parallel to get more insights into the dataset. 
 
The sum of daily scanned unique ports is an important indicator for attack detection. It can be seen that there will be a normal number of daily scanned unique ports in one network. Given no significant infrastructure changes, this normal number of daily scanned unique ports is usually stable across the whole network. For example, in our data analysis, this network has an observed number of 1000. However, when an attack occurs on certain days, the number of daily scanned unique ports can significantly increase. In our analysis, this number increases to 20\,000, given 66\,741 IP addresses in the network. Thus, we proposed to use a threshold-based approach in CopAS to consider a suspicious number of daily scanned unique ports.  

It is important to scale the attack indicator based on the normal number of daily scanned unique ports. That means that when the number of daily scanned unique ports deviates from the normal value, the extent of the deviation can be leveled to a different attack indication stage. For our analysis, it is easy to identify the abnormal behavior as an outlier number when around 20 times more than normal scans. However, we believe that in certain networks, the scan increase can also be caused by other events, such as installing new software or the scans that can be caused by the security software. Therefore, scaling the deviations from the normal number of daily scanned unique ports is an important step in approaching attack detection, such as insider attack detection. 

For each IP, there is a normal number of daily port scans. Therefore, ideally, all the IPs in the networks should maintain the normal level of daily port scans. When the number of daily port scans for one IP is significantly increased, there will be a possibility that this IP is launching an attack. Thus, this number can be tracked for each IP and can contribute to real-time attack detection. In our analysis, most IPs have 100 port scans each day. However, when the number is increased to 3000, there is a high possibility of an attack involving this IP. In a real-time setting, once the daily threshold of port scan is passed, the network may focus on monitoring the further behavior of this IP. 

\section{\uppercase{Conclusion}}
\label{sec6}

In this paper, we have introduced a forensic Big Data analytics platform called CopAS, a comprehensive and practically-usable solution for analyzing captured network traffic data at the level of individual network flows. Using a well-designed architecture and unique accompanying features, CopAS combines a suite of existing data analytics tools into a user-friendly environment that allows the data analyst to focus solely on the analysis itself rather than on building the necessary infrastructure and configuring the tools used.

As CopAS development and features are primarily inspired by its usability for real-world police investigators and their infrastructures, we employ its features based on their feedback. CopAS proves that it is a highly beneficial tool for day-to-day analyses, especially for smaller network captures analyzed on investigators' workstations, saving the centralized servers' resources. Besides this primary application, real-life CopAS usage has shown another suitable use case by serving as an easy-to-use training tool for new or inexperienced police data analysts, allowing them to familiarize themselves with the integrated analytical tools and examinations of various types of attacks before dealing with real-life datasets.

Beyond introducing CopAS architecture, this paper has demonstrated its usability in the case of detecting network-based insider attacks. We have employed the CopAS platform with real-world settings and experimented with a real-like PCAP dataset. The experimental results have identified the intrusions in the PCAP network captures. Further, we have examined that the CopAS system can be easily deployed in a cloud computing environment and deal with different data structures with reasonable run time. Thus, it can indicate which days the possible attacks might have occurred, help network administrators trace the possible IP(s) that launched the attacks, and monitor the port scan behaviors.

\section*{Acknowledgment}

The work was supported from ERDF/ESF ``CyberSecurity, CyberCrime and Critical Information Infrastructures Center of Excellence`` (No. CZ.02.1.01/0.0/0.0/16\_019/0000822).

\bibliographystyle{ACM-Reference-Format}
\input{main.bbl}

\balance
\end{document}